\documentclass[nonacm,sigplan]{acmart}
\usepackage{hyperref}
\newcommand{\sjm}{\textsc{SJMalloc}}
\newcommand{\bench}[1]{\multirow{3}*{\parbox{1.5cm}{#1}}}
\usepackage[normalem]{ulem}
\usepackage{multirow}
\usepackage{subcaption}
\usepackage{tikz}
\usetikzlibrary{chains,shapes,shapes.multipart, calc}
\usepackage{enumitem}
\begin{document}
\author{Stephan Bauroth, TU Berlin}
\email{s.bauroth@tu-berlin.de}

\title{\sjm{}: the security-conscious, fast, thread-safe and \\memory-efficient heap allocator}

\begin{abstract}
Heap-based exploits that leverage memory management errors continue to pose
a significant threat to application security. The root cause of these
vulnerabilities are the memory management errors within the applications,
however various hardened allocator designs have been proposed as mitigation.
A common feature of these designs is the strategic decision to store heap
metadata separately from the application data in use, thereby reducing the
risk of metadata corruption leading to security breaches.\\
Despite their potential benefits, hardened allocators have not been widely
adopted in real-world applications. The primary barrier to their adoption
is the performance overheads they introduce. These overheads can negatively
impact the efficiency and speed of applications, which is a critical
consideration for developers and system administrators.\\
Having learned from previous implementations, we developed
\sjm{}, a general-purpose, high-performance allocator that addresses these
concerns. \sjm{} stores its metadata out-of-band, away from the application's
data on the heap. This design choice not only enhances security but also
improves performance. Across a variety of real-world workloads, \sjm{}
demonstrates a $\sim6\%$ performance improvement compared to \texttt{GLibc}s
allocator, while using only $\sim5\%$ more memory.\\
Furthermore, \sjm{} successfully passes the generic elements of the
\texttt{GLibc} malloc testsuite and can thus be used as a drop-in replacement
for the standard allocator, offering an easy upgrade path for enhanced security
and performance without requiring changes to existing applications.
\end{abstract}

\maketitle
\pagestyle{plain}

\section{Introduction}
Heap-based attacks are increasingly gaining popularity and recognition
in the field of software security. This trend can be attributed to the
fact that other classes of attacks have been, and continue to be, mitigated
by the development and implementation of new defense mechanisms. Notable
examples of these mechanisms include stack canaries\cite{canary}, which
prevent stack-based buffer overflow attacks by detecting corruption of
stack variables, non-executable page protection, which restricts execution
rights to specific areas of memory, Address Space Layout Randomization
(ASLR)\cite{aslr}, which randomizes memory addresses used by system and
application processes, position-independent executables, which allow code
to execute regardless of its absolute address, and even hardware pointer
sanitization\cite{arm:ptrauth}, which ensures pointers are valid before
they are dereferenced.

While the prerequisites and execution details of heap-based attack vectors
vary significantly, they share a common feature: they exploit application's
memory management errors. These errors are combined with the internal
organization of the allocator’s data structures, which enables an attacker
to transform a simple memory management mistake into a viable exploit.
This commonality underscores the critical nature of robust memory management
practices and the need for continued research and development in this area.

Numerous allocators incorporating out-of-band metadata have been proposed
in the academic literature \cite{FreeGuard, DieHarder, cling}.
These allocators aim to mitigate the risk of heap-based attacks by isolating
metadata from application data. However, despite their theoretical advantages,
these allocators have not been widely adopted in practical systems. The primary
reason for this reluctance appears to be the performance overheads introduced
by these allocators.

In response to these challenges, we introduce \sjm{}, a novel allocator
designed to be resistant to the most common memory management errors. The
key innovation of \sjm{} lies in its approach to internal data structure
placement. Unlike traditional allocators, \sjm{} ensures that its internal
data structures do not reside in close proximity to an application's data on
the heap. Instead, these structures are placed separately (and possibly
sparsely) within the application's virtual address space. This design choice
significantly reduces the risk of memory management errors leading to
security vulnerabilities.

Moreover, \sjm{} is distinguished from other out-of-band allocators by
its efficient use of memory and its performance characteristics. Despite
incorporating out-of-band metadata, \sjm{} uses only marginally more memory
than the standard allocator provided by Glibc ($\sim5\%$). Additionally, it
exhibits negative performance overhead on average ($\sim6\%$), meaning that
it is faster than the standard allocator. This combination of security and
efficiency makes \sjm{} a compelling choice for applications where both
performance and security are paramount.

\paragraph{Outline}
The remainder of this paper is organized as follows. Section \ref{sec:background}
systemizes heap attacks. Section \ref{sec:terms} defines some terms before we formalize
some requirements for our allocator in Section \ref{sec:design}. In Section
\ref{sec:impl} we describe the actual implementation of \sjm{}. Section
\ref{sec:evaluation} evaluates \sjm{} considering the design goals and
performance characteristics as well as drawbacks. Finally, Section
\ref{sec:future} sets an outline for future work and possible improvements
to \sjm{}.

\section{Background}\label{sec:background}
Heap-based attacks continue to pose a significant threat to application security.
A thorough formal analysis of this issue can be found in the work of Novark
and Berger \cite{DieHarder}.\\
While we concur with their detailed examination of the nature and mechanisms
of heap-based exploits, we diverge in our perspective on which specific attacks
an allocator should be designed to defend against. To provide clarity and
structure to our discussion, we build on their analysis and categorize
heap-based exploits into two primary groups based on their objectives:
\begin{enumerate}
	\item \textbf{Attacking the allocator}: These exploits aim to manipulate
			the allocator into performing a write operation under the attacker's
			control.
	\item \textbf{Attacking the application}: These exploits seek to alter
			an application's intended behavior by exploiting concurrent uses
			of a specific block on the heap.
\end{enumerate}
We will explore both scenarios in more detail.

\paragraph{Arbitrary write by the allocator}
Exploits that compel the allocator to perform an arbitrary write for the
attacker exploit the fact that the allocator maintains its metadata within
the heap. These attacks typically occur when the allocator attempts to update
its metadata while operating on either overwritten metadata, as in the case
of a buffer overflow, or falsified metadata, such as through an invalid
\texttt{free()}).

By keeping the allocator's metadata outside of the heap, the burden on the
attacker increases significantly. In such a scenario, the attacker would
need to perform an arbitrary write directly into the allocator's external
metadata to be able to trigger an arbitrary write by the allocator. By
requiring an arbitrary write to trigger an arbitrary write, the attack is
useless. Consequently, fully segregated metadata effectively mitigates this
kind of heap-based attacks.

\paragraph{Concurrent uses of a given block}
Concurrent use of a given memory block is another critical attack surface
that attackers can exploit by manipulating the application to use the same
block of memory for different purposes simultaneously. This can occur
through several mechanisms:
\begin{enumerate}
	\item\textbf{Double Free}: This situation arises when the application
	mistakenly calls \texttt{free()} on the same memory block more than once.
	This can cause the allocator to return the same memory block to the
	application multiple times, leading to undefined behavior and potential
	security vulnerabilities.
	\item\textbf{Use After Free}: In this scenario, the application continues
	to use a memory block after it has been freed. If the allocator already
	reused the block to serve an allocation request, a conflict of usage
	is created, potentially leading to exploitation.
	\item\textbf{Buffer Overflow}: Here, the contents of a block of memory
	are altered by writing beyond its allocated boundaries. This can corrupt
	adjacent memory before the application accesses the corrupted data.
\end{enumerate}

A prominent example of exploitation through concurrent use is \texttt{vTable}-based
attacks against \texttt{C++} applications, where an attacker overwrites the
virtual table pointer of an object to execute arbitrary code\cite{rix}.\\
To mitigate these threats, an allocator can implement defenses such as:
\begin{enumerate}
    \item \textbf{Preventing Block Reuse}: By ensuring that a memory block
    is not returned to the application more than once, allocators can avoid
    double-free vulnerabilities. This implies an allocator always has to
    actually track the status of every memory block and not delay
    \texttt{free()} operations.
\end{enumerate}
However, other forms of concurrent memory block use, such as use-after-free
and buffer overflows, are not as easily defendable by the allocator alone.
These exploits do not involve the allocator directly, making it challenging
to address them solely through allocation strategies.\\
Further countermeasures have been proposed to enhance security, including:
\begin{enumerate}[resume]
	\item \textbf{Heap Canaries}: Placing certain special values (canaries)
	around allocated memory blocks can detect buffer overflows.
	\item \textbf{Delayed and Randomized Reuse of Blocks}: Reducing the
	predictability of memory reuse through introduced randomization makes
	exploitation more difficult.
\end{enumerate}

However, these methods introduce performance overhead and offer only statistical
improvements in security without guaranteeing detection of all attacks.
In conclusion, while allocators can implement specific defenses against
double-free vulnerabilities, comprehensive protection against all concurrent
use exploits requires additional strategies beyond the allocator itself.
These strategies can enhance security but often come with trade-offs in performance.

\section{Terminology}\label{sec:terms}
To aid the reader, we start with a few definitions of interest:
\begin{description}
	\item[Allocation] An allocation is a request by a program to be given
					a certain amount of memory. This request is issued via a
					defined API, typically a call to \texttt{malloc()}
					or \texttt{new}.
	\item[Cell]		A cell is a management structure representing the smallest
					unit of memory in a given size class. Cells contain
					metadata for allocations. They reside strictly
					outside the heap.
	\item[Block]	A block is a memory range on the heap that is either
					in use by an application or is free, i.e. not in use
					by the application. A block spans the memory represented
					by at least one cell (in variable bins, see Section
					\ref{sec:vbins}) or exactly one cell (in fixed bins,
					see Section \ref{sec:fbins}).
	\item[Bin]		A bin is a collection of cells.
\end{description}

\section{Design}\label{sec:design}
Learning from earlier implementations, we define four design goals:
\begin{itemize}
	\item	\textbf{avoid costly synchronization}: we want to achieve thread
			safety with as few locks and/or atomic operations as possible
	\item	\textbf{only one contiguous memory range}: metadata can
			not be assumed to be placed (virtually) contiguous in memory
	\item	\textbf{as few caches as possible}: to reliably detect double-frees,
			we want to employ caches only where absolutely necessary
	\item	\textbf{reasonable size classes}: to allow for sensible optimizations,
			we want to split allocations into categories by their respective size
\end{itemize}

\paragraph{Thread Synchronization}
We differentiate between the thread that allocated a block (the \textit{owning thread})
and those that did not (the \textit{remote threads}). This distinction allows us to
optimize performance by minimizing the use of locks and atomic operations.

To achieve this, we ensure that a block freed by the owning thread can operate
directly on its data structures without requiring locks or atomic operations.
This is accomplished by duplicating the relevant data structures: one set is
accessed exclusively by the owning thread, while the other set is used only
when a lock is held.

By segregating access in this manner, we allow the owning thread to manage
its allocated blocks efficiently, avoiding the overhead associated with
synchronization mechanisms. Remote threads, on the other hand, interact
with the duplicated data structures under controlled conditions, ensuring
thread safety without compromising performance. The implementation details
of this approach are elaborated in Section \ref{sec:cross}.

\paragraph{No second growing memory range}
We do not assume that the operating system can place a second growing memory
range within the virtual address space. Therefore, we must develop an efficient
method for associating a given pointer with its corresponding metadata location,
a component we call \textit{reverse lookup.}\\
The detailed implementation of this reverse lookup mechanism is provided
in Section \ref{sec:reverse}, where we discuss the data structures and
routines employed to achieve this association efficiently and reliably.

\paragraph{No caching}
We do not employ any form of caching for recently freed blocks\footnote{except
for external allocations, see Section \ref{sec:external}}. To reliably
detect double frees, it is essential to check the actual status of each
memory block. If caching were used, the allocator would need to verify the
entire cache to ensure that a given block is not already present, thereby
negating any potential performance benefits.

By avoiding caching, we streamline the process of checking the status of
memory blocks, ensuring that double frees are accurately detected. This
detection is crucial for preventing exploits that rely on the allocator
returning the same memory block multiple times after it has been freed
more than once.

Merely detecting a double free allows the allocator to neutralize any potential
exploitability associated with this vulnerability. Double \texttt{free()}
exploits depend on the allocator's reuse of the same block, which can lead
to security breaches if not properly managed. By eliminating the possibility
of reusing a double freed block, we enhance the overall security and
robustness of the allocation system.

\paragraph{Size classes}
To apply reasonable optimizations and make the best trade-offs, we categorize
allocations into three size classes:
\begin{enumerate}
	\item \textbf{Small Allocations (served by fixed bins)}\\
			Allocations up to \texttt{FBIN\_MAX\_SIZE} (512 bytes) are
			serviced using \textit{fixed bins} (\texttt{FBIN}s). These bins
			operate in a BiBoP (Big Bag of Pages) style, where blocks always
			maintain their size, ensuring efficient handling of small memory
			requests. Details of this mechanism are provided in \ref{sec:fbins}.
	\item \textbf{Medium Allocations (served by variable bins)}\\
			Allocations above \texttt{FBIN\_MAX\_SIZE} but below\\
			\texttt{MMAP\_THRESHOLD} (128 kB) are managed using
			\textit{variable bins} (\texttt{VBIN}s). This novel approach
			balances internal fragmentation within the heap with the overhead
			of bookkeeping. \texttt{VBIN}s offer a flexible and efficient
			way to handle medium-sized allocations. Further details can be
			found in Section \ref{sec:vbins}.
	\item \textbf{Large Allocations (forwarded to \texttt{mmap()})}\\
			Allocations exceeding \texttt{MMAP\_THRESHOLD} are directly forwarded
			to \texttt{mmap()}. This strategy prevents fragmentation within
			the heap, a common practice among allocators to efficiently
			manage large memory requests. The specifics of this mechanism
			are explained in Section \ref{sec:external}.
\end{enumerate}
By segmenting allocations into these three size classes, we optimize memory
management for various allocation sizes, achieving a balance between performance,
fragmentation, and overhead. Each class is tailored to handle specific
allocation ranges effectively, ensuring that memory is utilized efficiently
across different use cases.

\section{Implementation}\label{sec:impl}
\sjm{}{} is implemented in pure C, designed with minimal dependencies, and
intended to be a secure drop-in replacement for \textsc{Glibc}'s
\textsc{ptmalloc}.

\subsection{Bins}\label{sec:bins}
Each bin in our allocator holds 1024 cells of uniform size. The metadata
of each bin includes the following information:
\begin{itemize}
	\item	\textbf{\texttt{shift}}:		Indicates the cell size within the bin.
	\item	\textbf{\texttt{next}}:			Pointer used to link bins into lists.
	\item	\textbf{\texttt{base}}:			Pointer marking the beginning of the
											managed heap area for the bin.
	\item	\textbf{\texttt{thread\_id}}:	Identifies the owning thread of the bin.
	\item	\textbf{\texttt{type}}:			Specifies whether the bin is a fixed
											bin or a variable bin.
\end{itemize}
To enable operations by other threads on cells within a bin, additional fields include:

\begin{itemize}
	\item	\textbf{\texttt{remote\_free\_count}}:
			Tracks the number of cells freed remotely.
	\item	\textbf{\texttt{remote\_freed}}:
			Bitmask indicating which cells have been remotely freed.
\end{itemize}
Both \texttt{FBIN}s and \texttt{VBIN}s share more fields, but their usage
differs between the two types, as elaborated in the following subsections.

\subsection{Fixed bins (\texttt{FBIN}s)}\label{sec:fbins}
Within FBINs, each cell corresponds to a fixed-size block. These blocks are
never merged or resized. Allocations within the size range of FBINs are
rounded up to the next power of two, which introduces some internal
fragmentation but provides significant performance benefits\footnote{In
contrast to claims in \cite{FreeGuard}, where performance penalties were
assumed for BiBoP-style allocators}. Per-block metadata is limited to a
single bit - the blocks current availability status (free or used). This
state is encoded in a bitmap residing directly behind the bin structure.
The bin’s metadata includes:
\begin{itemize}
	\item	\textbf{\texttt{free\_cnt}}: Number of currently free blocks.
	\item	\textbf{\texttt{free\_head}}: Points to the first word of the
				bitmap where a bit is set to indicate a free block.
	\item	\textbf{\texttt{free\_bits}}: The bitmap recording the state of
				all\\ blocks within this bin.
\end{itemize}
To optimize the handling of bitmaps, we use \texttt{uint\_fast16\_t} as the
data type, allowing the compiler to select the most efficient integer type
supported by the platform, speeding up scanning for free blocks.
Typical operations on the bitmask translate directly to compiler builtins
or logical operations, most notably \texttt{find first set}, \texttt{OR}
and \texttt{AND}.

\subsection{Variable bins (\texttt{VBIN}s)}\label{sec:vbins}
To mitigate the internal fragmentation introduced by \texttt{FBIN}s,
especially as allocation sizes increase, VBINs adopt a strategy of rounding
allocation sizes up to multiples of 16 rather than powers of 2. However, a
purely bitmask-based approach is insufficient for accurately representing
these allocations. In \texttt{VBIN}s, each cell corresponds to a specific
amount of memory on the heap, determined by the bin's shift value. Within
each \texttt{VBIN}, two double-linked lists are maintained:
\begin{itemize}
	\item \textbf{Free List}: Orders all free blocks by size.
	\item \textbf{Spatial List}: Orders all blocks by the index of their starting cell.
\end{itemize}
Allocations within a \texttt{VBIN} are always larger than the size of a cell.
For Example, the smallest allocation handled by a \texttt{VBIN} with cell size
of 512 bytes is 528 bytes. This design ensures that each cell's memory range
on the heap can accommodate the start of at most one allocation.

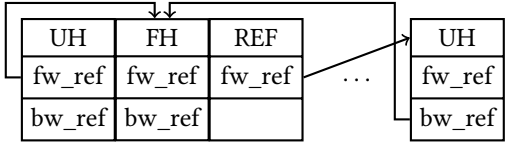
\begin{figure}
	\centering
	\begin{tikzpicture}[cell/.style={
		rectangle split,
		rectangle split parts=#1,
		draw,
		minimum height=5cm}]
		\tikzstyle{every path}=[thick]
		\begin{scope}[start chain=1 going right,node distance=-0.15mm]
			\node [on chain=1, cell=3] (uh1) {UH \nodepart{two}fw\_ref \nodepart{three} bw\_ref};
			\node [on chain=1, cell=3] (fh1) {FH \nodepart{two} fw\_ref \nodepart{three} bw\_ref};
			\node [on chain=1, cell=3] (ref) {REF \nodepart{two} fw\_ref \nodepart{three} \color{white}{bw\_ref}};
			\node [on chain=1, cell=1, draw=none] {};
			\node [on chain=1, cell=1, draw=none] (space1) {$\ldots$};
			\node [on chain=1, cell=1, draw=none] {};
			\node [on chain=1, cell=3] (uh2) {UH \nodepart{two}fw\_ref \nodepart{three} bw\_ref};
		\end{scope}
		\draw[->] (ref.two east) -- (uh2.text west);
		\draw[->] (uh1.two west) -| ($(uh1.north west)+(-.2,.2)$) -| ($(fh1.north)+(-.1,0)$);
		\draw[->] (uh2.three west) -| ($(uh2.north west)+(-.2,.2)$) -| ($(fh1.north)+(.1,0)$);
	\end{tikzpicture}
	\caption{Spatial list composition in free heads: the previous cell is
			a used head, thus removing the necessity to store its index,
			the next cell is not, making space to store the next used heads
			index.}\label{fig:free_heads}
\end{figure}

Each cell within a VBIN contains information about the block starting in
its memory range, if applicable. Cells are classified into four types:
\begin{itemize}
	\item	used head (\textbf{UH}):\\
			A currently allocated block starts within this cell
	\item	free head (\textbf{FH}):\\
			A currently free block starts within this cell
	\item	reference (\textbf{REF}):\\
			This cell contains data for one of its neighbours
	\item	unused (\textbf{UN}):\\
			This cell is unused (i.e. somewhere in the middle of a block)
\end{itemize}
Each cell is 32 bits in size, its type is encoded in 2 of those bits.
Depending on the type, the remaining bits serve different purposes:
\paragraph{\textbf{Used Head (UH)}}
Each used head contains \texttt{fw\_ref} and \texttt{bw\_ref} (10 bits each),
which store the forward and backward references for the spatial list,
respectively. These references are indices of the next and previous cells
in the list. The remaining 10 bits are used to store the offset within the
memory that this cell represents, indicating where the block starts within
that memory range.
\paragraph{\textbf{Free head (FH)}}
Free heads are part of two distinct double-linked lists: the free list and
the spatial list. Since the \texttt{fw\_ref} and \texttt{bw\_ref} fields
alone are not sufficient to store all the required information, we use them
differently compared to used heads. They form the size-ordered free list,
which helps in quickly finding free blocks of the required size. To form
the regular spatial list of all heads, we handle two cases:
\begin{itemize}
	\item If the next (or previous) cell contains a used head, it directly
			provides the correct reference.
	\item If the next (or previous) cell does not contain a used head, we
			use it to store the correct reference instead. We call these
			cells \textit{references} (REFs).
\end{itemize}
A visualization of both cases is provided in Figure \ref{fig:free_heads}.
The \texttt{offset} field in free heads is used in the same way as in used
heads, indicating the starting point of the free block within the memory
range this cell represents.

\paragraph{\textbf{Reference (REF)}}
These cells only contain a reference to the next or previous used head. They
store this reference in \texttt{fw\_ref} or \texttt{bw\_ref}, respectively.
Additionally, the two lowest bits of the offset field are used to distinguish
forward from backward references, which aids in debugging and sanity checks.
Since no two free heads can ever be adjacently located (they would
be merged if that occurred), a REF cell can only be valid in one direction.
This fact allows us to mark a special cell within each bin: the last cell,
which can never be a head (neither free nor used). Therefore, this last
cell is used to store the head and tail of the free list in \texttt{fw\_ref}
and \texttt{bw\_ref}. It is marked as a reference in both directions.

\paragraph{unused (UN)}
These cells are marked as unused, with all fields except the type expected to be zero.

\vspace{.1cm}
When allocating memory, we search for the first bin that contains a block
large enough to fulfill the request. By checking the size of the first cell
in the free list, we can determine if the bin holds a suitable block. Once
a suitable bin is found, we use the double-linked free list within that bin
to implement a best-fit approach. This approach minimizes the need for splitting
and merging operations, enhancing efficiency.

\texttt{VBINs} employ a few fields of their bin structure differently from
how \texttt{FBIN}s use them:
\begin{itemize}
	\item	\textbf{\texttt{free\_head}}: Unused, the actual free head is
				stored in the last cell together with the tail of the free
				list.
	\item	\textbf{\texttt{free\_cnt}}: Holds a secondary free head that
				points to a cell within the lower half of the free list --
				so we can save on iterations if looking for smaller blocks
	\item	\textbf{\texttt{cells}}: Array of this bins cell structures
\end{itemize}

\begin{figure*}
	\centering
	\includegraphics[width=.7\linewidth]{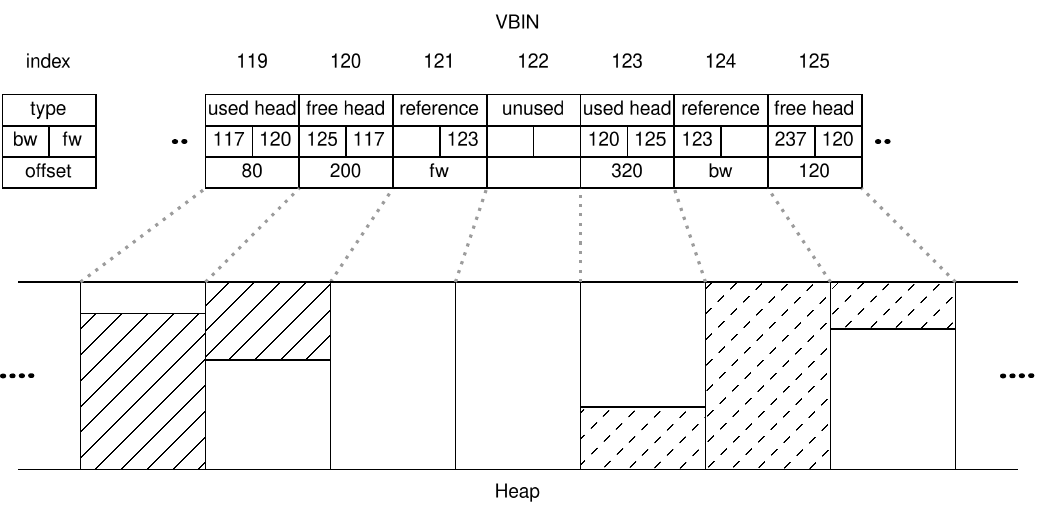}
	\caption{A portion of a \texttt{VBIN} and its associated heap memory.
	The \texttt{VBIN}s cells are shown at the top, the heap memory at the bottom.
	Striped memory denotes used memory, white memory is free. Each cell contains
	the metadata of the memory block \textbf{starting} in the cell's associated
	memory. Note the forward reference in cell 121 for the free head in cell 120
	and the backward reference in cell 124 for the free head in cell 125 - while
	cell 119 does not contain a backwards reference for the free head in cell 120
	because it already is the used head that would be referenced.}\label{fig:vbin}
\end{figure*}

A breakdown of a hypothetical \texttt{VBIN} and its associated heap space
is given in Figure \ref{fig:vbin}.

\subsection{External Allocations}\label{sec:external}
To avoid fragmentation within the heap, large allocations are directly
forwarded to \texttt{mmap()}. Because both \texttt{munmap()} and \texttt{mremap()}
require valid \texttt{len} or \texttt{old\_size} arguments, we store the size
of each mapping in a data structure referred to as the \textit{external lookup.}
The external lookup is an array of structures that track the current state
of each mapping. Each structure contains the following information:
\begin{itemize}
	\item \textbf{Base Address}: The starting address of the memory region mapped.
	\item \textbf{Size}: The size of the mapping.
	\item \textbf{State}: Indicates whether the mapping is currently in use or free.
	\item \textbf{Free List References}: Two indices that form a double-linked list
					of freed mappings, ordered by size, that are available for reuse.
\end{itemize}
To optimize performance, we maintain a minimal cache of external mappings
to reduce the number of system calls. When inserting a new element into the
list, we ensure that it is not already present. Unlike for smaller allocations,
this approach enhances performance due to the relatively high cost of system calls.

\subsection{Reverse Lookup (revlookup)}\label{sec:reverse}
When a call to \texttt{free()} is made, the allocator needs to locate the
metadata corresponding to the pointer that was passed to it. In-band allocators
achieve this by subtracting a fixed offset to find the beginning of the block's
header. However, this method does not work for out-of-band allocators. Instead,
a data structure is required to map a pointer to its associated metadata.

Some out-of-band allocators store their metadata at the beginning of each
chunk, bin, or arena (e.g., \texttt{jemalloc} \cite{jemalloc}), making it
more challenging but still possible to overwrite. Other out-of-band allocators,
like \texttt{FreeGuard} \cite{FreeGuard}, assume a virtually contiguous
memory region is available for their metadata, allowing for a linear mapping
between the heap and metadata. In contrast, we, like \texttt{DieHarder}
\cite{DieHarder}, do not base our reverse lookup on such assumptions or
place metadata near the heap. \texttt{DieHarder} uses a hash map to find
a pointer's metadata.

Our approach involves subtracting the heap’s start address from the given
pointer and dividing the result by the minimum bin size, which provides an
index of that bin on the heap\footnote{Given that bins manage cells of
different shifts, larger bins cover multiple indices}. We use this index
to access a tree-like structure that contains the addresses of the bin’s
metadata. This tree-like structure is inspired by multi-level page tables,
consisting of a hierarchical collection of tables pointing to the next table's
location until a target depth is reached. At this depth, the table contains
the address of the sought metadata.

The location of the reverse lookup structure is stored in a process-global
pointer. This pointer always contains the address of a page, which includes
trailing zeros. We use these free bits to encode the current depth of the
revlookup. To find the actual metadata for a given pointer, we mask out
the depth and dereference the resulting pointer, offsetting with the relevant
part of our index, and then decrement the depth. When the depth reaches zero,
we obtain the address of the actual metadata.

\subsection{Cross-thread \texttt{free()}s}\label{sec:cross}
When a block is freed by a different thread than the one that originally
allocated it, we detect this by using the \texttt{thread\_id} stored in
the bin's management data. Instead of directly modifying the data structures
that represent the bin's state, we mark the cell that starts this block
(for \texttt{VBIN}s), or actually is this block (for \texttt{FBIN}s), as
''remotely freed.''

After identifying a remotely freed block, we grab the bin’s \texttt{mutex}
to ensure thread safety. We then increment the bin’s \texttt{remote\_free\_cnt}
and set the corresponding bit in \texttt{remote\_freed}, which keeps track
of which cells have been freed by remote threads.

Whenever an owning thread (the thread that originally allocated the memory)
frees a cell in any of its bins, it checks the bin's \texttt{remote\_free\_cnt}.
If this count is non-zero, it indicates that there are cells marked as
remotely freed. The owning thread then locks the bin’s \texttt{mutex}
to safely access and modify the bin's data structures. It proceeds to
actually free all cells that are marked as remotely freed, updating the
bin’s state accordingly.

\section{Evaluation}\label{sec:evaluation}
\subsection{Security}
In this section, we will evaluate the security of \sjm{}.
\paragraph{Threat Model}
We assume an attacker that is able to trigger memory management errors in
an application at her will and in arbitrary order. The goal of the attacker
is to form a chain of memory management errors to trigger an arbitrary
write by the allocator. We consider double frees, invalid frees and buffer
over- or underflows to be valid primitives to attack an allocator. Use-after-frees
(excluding double frees) do not attack the allocator but the application
itself.\\
Our threat model excludes attackers who can already trigger arbitrary writes,
as exploiting the allocator in such cases would be redundant. We also do not
consider exploit chains that lead to data-driven attacks against the application
itself, as we consider defending against these within the allocator infeasible.

\paragraph{Double Free}
Setting aside external allocations, \sjm{} does not use caches, so double frees are always detected.
However, if a remote thread performs the second free after a native thread's
first free, detection is delayed until the owning thread attempts to free
the block. Multiple remote frees will be detected immediately.
For external allocations, the complete cache is checked, and thus the double free is detected.
Regardless of detection timing, a double free will never result in \sjm{} issuing
the same block of memory to two different requests, thus preventing exploitation.

The consequences of a detected double free can be configured at compile
time: ignore, report or report and crash.

\paragraph{Invalid Free}
An invalid free that does not constitute a double free (addressed above)
is reliably detected across all three size classes. Similar to double frees,
the allocator's behavior upon detecting an invalid free (ignore, report, crash)
can be configured at compile time.

\paragraph{Buffer Overflows and Underflows}
With external metadata not co-located with the application's data, buffer
overflows and underflows do not affect the metadata but rather other data
in use by the application. While this still provides potential exploitation
vectors (data-driven attacks), it removes those targeting the allocator's
implementation.\\
Defense against data-driven attacks can be aided in an allocator, by routinely
checking canary values placed on the heap and other means, but every
method proposed so far decreases performance significantly, hindering adoption
in the ecosystem.

Conclusively, \sjm{} effectively removes all exploit vectors that could cause
the allocator to execute an arbitrary write on behalf of the attacker. However,
the application itself is not defended by \sjm{}, as providing such protection
would compromise performance benefits.

\subsection{Performance}
We evaluate \sjm{} using mimalloc-bench \cite{mimalloc-bench}, an open source
allocator benchmarking suite developed to benchmark mimalloc \cite{mimalloc},
an alternative heap allocator developed and open sourced by Microsoft. We
include pull request 194\footnote{\url{https://github.com/daanx/mimalloc-bench/pull/194}},
which adds compiling Linux as a benchmark to extend the number of real world
workloads benchmarked.

\begin{figure*}
	\centering
	\begin{subfigure}{.48\linewidth}
		\centering
		\includegraphics[width=.99\linewidth]{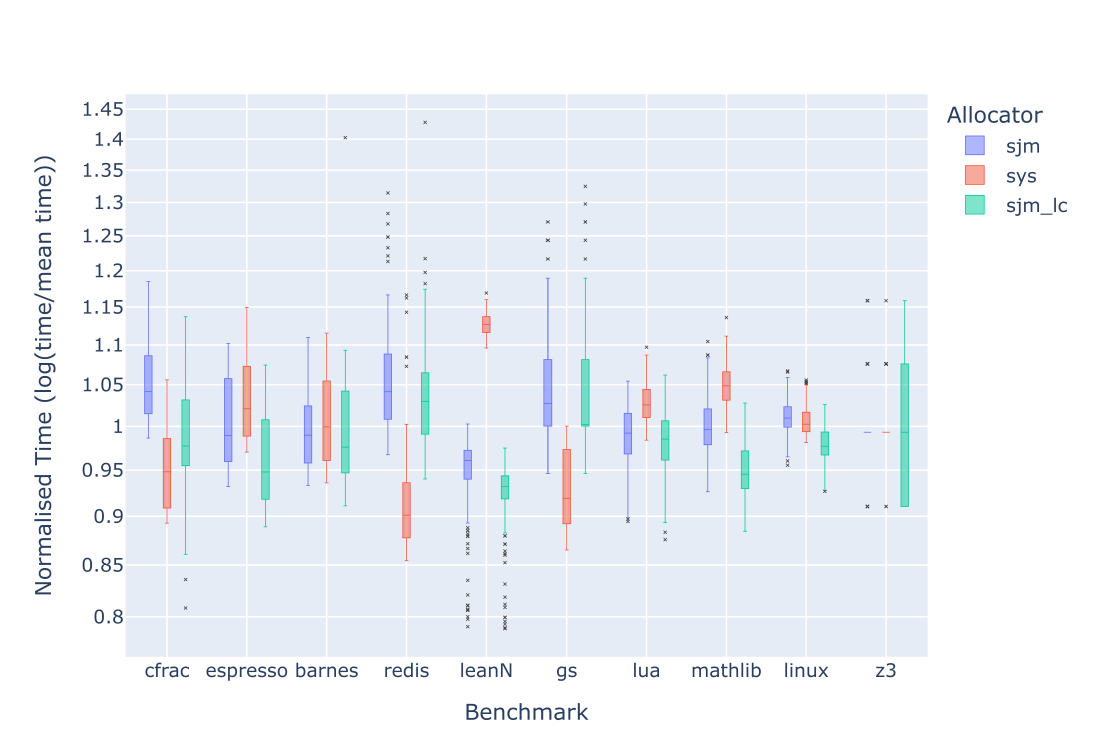}
	\end{subfigure}
	\hfill
	\begin{subfigure}{.48\linewidth}
		\centering
		\includegraphics[width=.99\linewidth]{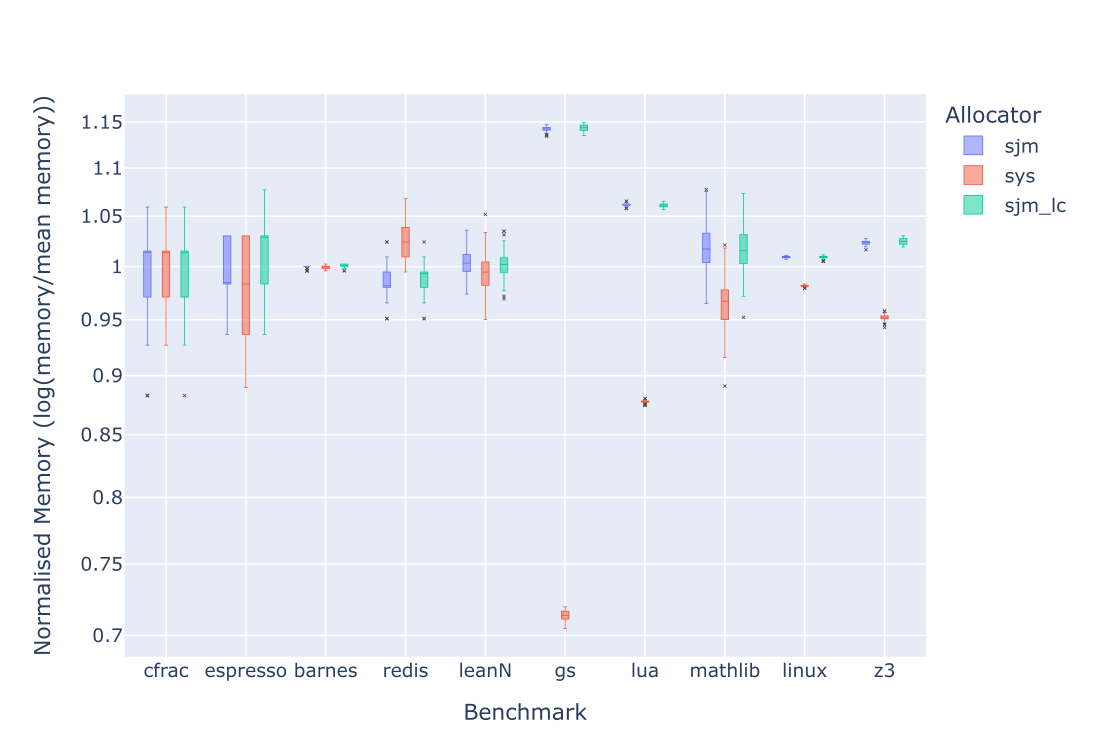}
	\end{subfigure}
	\caption{Normalized runtime (left) and normalized memory usage (right)
	of all real-world benchmarks in \texttt{mimalloc-bench}. In both cases,
	lower is better.}\label{fig:results}
\end{figure*}

We benchmark two versions of \sjm{}: one built isolated and employed using
\texttt{LD\_PRELOAD} (\texttt{sjm}) and one built into \texttt{glibc} 2.39
(\texttt{sjm\_lc}) instead of its standard allocator (\texttt{ldmalloc}).
\texttt{sys} is the baseline, \texttt{ldmalloc} from \texttt{Glibc} 2.39.
The results of all benchmarks are shown in Table \ref{tab:results}. The
results of all real-world workloads are shown in Figure \ref{fig:results}.

The suite is run 10 times (\texttt{-r=10}), each individual benchmark is
run 20 times (\texttt{-n=20}) resulting in 200 data points per benchmark
and allocator. Across the complete benchmark suite, \texttt{sjm\_lc} exhibits
a runtime overhead of 5.45\% while using 26.05\% less memory. Across the set
of real-world workloads included in the suite, \texttt{sjm\_lc} is 5.76\%
faster while using 5.02\% more memory. Three of the benchmarks show results
that are worth discussing.

\textbf{\texttt{larsonN(-sized)}}\cite{larson}\\
This benchmark was designed to mimic the behaviour of long-running
applications such as web or mail servers. Each worker thread has a given
number of prepared allocations to work with. It frees a randomly chosen
block of said set and re-allocates it with a random size (within a
given range). After a given number of iterations over the array of blocks,
the thread spawns a copy of itself and terminates -- leaving a few allocated
blocks behind for the newly spawned thread to free. The benchmark is stopped
after a fixed amount of time, completed operations are counted and a
''relative time'' is computed.\\
\sjm{} performs significantly slower than \texttt{ldmalloc} in this benchmark
due to the random order of operations and varying sizes, which negatively
affect its internal structures: both the free list of bins that each thread
keeps as well as the internal free lists of \texttt{VBIN}s grow very large.
This slows down iterations over these lists.\\
We consider it noteworthy that none of the real-world workloads exhibited
such behaviour.

\textbf{gs}\\
The ghostscript benchmark converts a single PDF file to ghostscript. This
benchmark heavily relies on large external allocations -- and \sjm{}'s
cache for external allocations is not as effective as \texttt{ldmalloc}'s.

\textbf{mstressN}\\
Similar to \texttt{larson}, \texttt{mstress} uses mutliple threads to
randomly re-allocate blocks of different sizes. It is not designed as a
benchmark, as stated in the README:\\
''This is a stress test for allocators, using multiple threads and retaining
some objects, re-allocating, and transferring objects between threads. This
is not a typical program workload but uses a random size distribution -- do
not use this test as a benchmark!''\\
\texttt{ldmalloc} uses a lot more memory (892 MB, 73.5\%) than \sjm{} in
this stress test, which can only be explained with some corner case in the
implementation that is present in \texttt{Glibc} 2.39.

\vspace{.1cm}
Much like Evans in \cite{jemalloc}, we prioritize an allocator's performance
in real-world applications as the most important metric,
leaving synthetic workloads for profiling and completeness tests:
''The only definitive measures of allocator performance are attained by
measuring the execution time and memory usage of real applications.''
\cite[pg. 1]{jemalloc}

With \sjm{} being 5.76\% faster than the current \texttt{Glibc} standard
allocator for real-world workloads, we consider our performance goals achieved.

\begin{table*}
	\setlength{\tabcolsep}{0.14cm}
	\begin{tabular}{|p{1.5cm}|r||rr|rr||p{1.5cm}|r||rr|rr|}\hline
		bench					& allocator	& \multicolumn{2}{c|}{runtime in sec}	& \multicolumn{2}{c||}{max. RSS in kB}	& bench						& allocator	& \multicolumn{2}{c|}{runtime in sec}	& \multicolumn{2}{c|}{max. RSS in kB}	\\\hline
		\bench{cfrac}			& sys		& 8.33		& 							& 2921.0	& 							& \bench{cache-scratch1}	& sys		& 1.81		& 							& 3697.3	& 							\\
								& sjm		& 9.23		& 10.80\%					& 2876.2	& -1.53\%					&							& sjm		& 1.91		& 5.86\%					& 3765.8	& 1.85\%					\\
								& sjm\_lc	& 8.71		& 4.55\%					& 2903.0	& -0.61\%					&							& sjm\_lc	& 1.84		& 2.07\%					& 3831.7	& 3.64\%					\\\hline
		\bench{espresso}		& sys		& 6.73		& 							& 2690.6	& 							& \bench{cache-scratchN}	& sys		& 0.61		&							& 3644.2	&							\\
								& sjm		& 6.54		& -2.82\%					& 2730.2	& 1.47\%					&							& sjm		& 0.58		& -4.13\%					& 3826.6	& 5.01\%					\\
								& sjm\_lc	& 6.30		& -6.48\%					& 2779.5	& 3.31\%					&							& sjm\_lc	& 0.58		& -4.39\%					& 3811.2	& 4.58\%					\\\hline
		\bench{barnes}			& sys		& 3.65		& 							& 58045.4	&							& \bench{glibc-simple}		& sys		& 5.39		&							& 1768.3	&							\\
								& sjm		& 3.61		& -0.97\%					& 58086.4	& 0.07\%					&							& sjm		& 4.83		& -10.33\%					& 1802.2	& 1.92\%					\\
								& sjm\_lc	& 3.61		& -1.12\%					& 58145.9	& 0.17\%					&							& sjm\_lc	& 4.28		& -20.50\%					& 1811.2	& 2.42\%					\\\hline
		\bench{redis}			& sys		& 4.67		&							& 8988.8	&							& \bench{glibc-thread}		& sys		& 7.51		&							& 2681.6	&							\\
								& sjm		& 5.40		& 15.47\%					& 8630.4	& -3.99\%					&							& sjm		& 10.16		& 35.25\%					& 2668.8	& -0.48\%					\\
								& sjm\_lc	& 5.31		& 13.64\%					& 8629.1	& -4.00\%					&							& sjm\_lc	& 9.07		& 20.67\%					& 2723.2	& 1.55\%					\\\hline
		\bench{leanN}			& sys		& 43.72		&							& 557472.0	&							& \bench{rocksdb}			& sys		& 6.88 		&							& 88835.3	&							\\
								& sjm		& 36.79		& -15.84\%					& 563135.4	& 1.02\%					&							& sjm		& 7.16		& 4.06\%					& 121163.0	& 36.39\%					\\
								& sjm\_lc	& 35.84		& -18.01\%					& 562263.6	& 0.86\%					&							& sjm\_lc	& 6.93		& 0.71\%					& 121177.4	& 36.41\%					\\\hline
		\bench{larsonN-sized}	& sys		& 29.84		& 							& 193401.7	&							& \bench{larsonN}			& sys		& 28.57		&							& 193103.9	&							\\
								& sjm		& 52.73		& 76.70\%					& 72776.3	& -62.37\%					&							& sjm		& 49.37		& 72.79\%					& 75127.0	& -61.10\%					\\
								& sjm\_lc	& 48.96		& 64.05\%					& 75491.2	& -60.97\%					&							& sjm\_lc	& 46.75		& 63.62\%					& 77299.8	& -59.97\%					\\\hline
		\bench{mstressN}		& sys		& 1.78		&							& 1213782.1	&							& \bench{mathlib}			& sys		& 3.00		&							& 126176.9	&							\\
								& sjm		& 1.33		& -25.31\%					& 321847.0	& -73.48\%					&							& sjm		& 2.86		& -4.65\%					& 133191.7	& 5.56\%					\\
								& sjm\_lc	& 1.28		& -28.07\%					& 321629.3	& -73.50\%					&							& sjm\_lc	& 2.72		& -9.41\%					& 133054.7	& 5.45\%					\\\hline
		\bench{rptestN}			& sys		& 2.41		&							& 32380.5	&							& \bench{linux}				& sys		& 110.93	&							& 173159.0	&							\\
								& sjm		& 2.00		& -17.06\%					& 27149.4	& -16.15\%					&							& sjm		& 111.57	& 0.58\%					& 178035.2	& 2.82\%					\\
								& sjm\_lc	& 1.96		& -18.60\%					& 26873.6	& -17.01\%					&							& sjm\_lc	& 107.94	&-2.70\%					& 178023.0	& 2.81\%					\\\hline
		\bench{gs}				& sys		& 0.34		&							& 30411.1	&							& \bench{malloc-large}		& sys		& 3.06		&							& 534205.7	&							\\
								& sjm		& 0.38		& 12.13\%					& 48637.7	& 59.93\%					&							& sjm		& 2.21		& -27.77\%					& 619295.4	& 15.93\%					\\
								& sjm\_lc	& 0.38		& 11.96\%					& 48707.6	& 60.16\%					&							& sjm\_lc	& 2.20		& -28.03\%					& 619284.5	& 15.93\%					\\\hline
		\bench{lua}				& sys		& 8.17		&							& 77826.8	&							& \bench{mleak10}			& sys		& 0.13		&							& 2576.6	&							\\
								& sjm		& 7.86		& -3.77\%					& 94106.9	& 20.92\%					&							& sjm		& 0.13		& -3.00\%					& 1833.0	& -28.86\%					\\
								& sjm\_lc	& 7.81		& -4.36\%					& 94092.2	& 20.90\%					&							& sjm\_lc	& 0.13		& -3.79\%					& 1875.2	& -27.22\%					\\\hline
		\bench{alloc-test1}		& sys		& 5.70		&							& 14018.6	&							& \bench{mleak100}			& sys		& 1.37		&							& 8892.8	&							\\
								& sjm		& 6.58		& 15.45\%					& 14092.2	& 0.53\%					&							& sjm		& 1.34		& -2.59\%					& 2015.4	& -77.34\%					\\
								& sjm\_lc	& 6.29		& 10.23\%					& 14001.3	& -0.12\%					&							& sjm\_lc	& 1.30		& -5.19\%					& 2040.3	& -77.06\%					\\\hline
		\bench{alloc-testN}		& sys		& 11.97		&							& 15752.3	&							& \bench{rbstress1}			& sys		& 5.41		&							& 129745.9	&							\\
								& sjm		& 14.43		& 20.63\%					& 14233.0	& -9.65\%					&							& sjm		& 5.35		& -1.05\%					& 113466.2	& -12.55\%					\\
								& sjm\_lc	& 13.47		& 12.54\%					& 14240.6	& -9.60\%					&							& sjm\_lc	& 5.27		& -2.66\%					& 112537.1	& -13.26\%					\\\hline
		\bench{sh6benchN}		& sys		& 2.86		&							& 424742.4	&							& \bench{rbstressN}			& sys		& 5.49		&							& 169111.1	&							\\
								& sjm		& 1.91		& -33.20\%					& 348942.1	& -17.85\%					&							& sjm		& 5.24		& -4.58\%					& 140901.8	& -16.68\%					\\
								& sjm\_lc	& 1.59		& -44.29\%					& 348389.1	& -17.98\%					&							& sjm\_lc	& 5.13		& -6.46\%					& 140833.9	& -16.72\%					\\\hline
		\bench{sh8benchN}		& sys		& 9.26		&							& 164663.0	&							& \bench{cache-thrash1}		& sys		& 1.75		&							& 3681.3	&							\\
								& sjm		& 5.89		& -36.40\%					& 129491.1	& -21.36\%					&							& sjm		& 1.78		& 1.48\%					& 3754.9	& 2.00\%					\\
								& sjm\_lc	& 4.93		& -46.69\%					& 129099.0	& -21.60\%					&							& sjm\_lc	& 1.76		& 0.49\%					& 3733.8	& 1.43\%					\\\hline
		\bench{xmalloc-testN}	& sys		& 4.64		&							& 54561.7	&							& \bench{cache-thrashN}		& sys		& 0.59		&							& 3653.8	&							\\
								& sjm		& 2.02		& -56.39\%					& 49467.2	& -9.34\%					&							& sjm		& 0.57		& -4.18\%					& 3835.5	& 4.97\%					\\
								& sjm\_lc	& 1.47		& -68.41\%					& 49447.0	& -9.37\%					&							& sjm\_lc	& 0.57		& -3.65\%					& 3822.1	& 4.61\%					\\\hline
								&			&			&							&			&							& \bench{z3}				& sys		& 0.12		&							& 44099.1	&							\\
								&			&			&							&			&							&							& sjm		& 0.12		& -0.98\%					& 47439.4	& 7.57\%					\\
								&			&			&							&			&							&							& sjm\_lc	& 0.12		& -2.13\%					& 47486.7	& 7.68\%					\\\hline
	\end{tabular}
	\caption{Runtime and memory usage of benchmarks that are part of \texttt{mimalloc-bench}. Each benchmark is run 20 times, the whole suite is repeated 10 times, resulting in 200 data points per allocator and benchmark}\label{tab:results}
\end{table*}

\subsection{Drawbacks}
Given the metadata \sjm{} needs to keep and that it is kept outside of the
heap, applications might run into situations where they exceeded their limit
for memory mappings(e.g., \texttt{vm.max\_map\_count} in Linux) faster than
when using a different allocator. See Section \ref{sec:future} for a proposed
solution.

\section{Future work}\label{sec:future}
\sjm{} does not defend against data-driven attacks, due to no reliable way
of detection. Further improvements in security on the heap could be achieved
when aiding software applications in detecting both the underlying memory
management issue as well as an ongoing attack.

Furthermore, an operating system could aide in improving the performance
of out-of-band allocators in two ways: guarantee a second growing memory
range in the virtual address space, rendering the reverse lookup down to
a linear lookup, and offering \texttt{mremap} and \texttt{munmap} in variants
that do not need to be given the \texttt{old\_size} of the mapping but
instead assume the whole mapping is to be dealt with. This would remove
the need for duplicated metadata within the allocator (\textit{external
lookup}).

Finally, \sjm{} might be tuned to varying tradeoffs. Most prominent,
metadata, the revlookup and the external lookup of \sjm{} currently
\texttt{mmap} memory from the operating system one page at a time. Asking
the operating system for larger chunks of memory would reduce performance
overhead (through fewer syscalls) and decrease the process' map count but
increase memory overhead -- which is why we leave this to users of \sjm{}.

\section{Acknowledgements}
To be disclosed later.

\section{Availability}
An anonymized verison of \sjm{} is available at \\
\url{https://anonymous.4open.science/r/sjmalloc-EFD9}.

\bibliographystyle{plain}
\bibliography{references}

\end{document}